\documentclass[letterpaper,twocolumn,10pt]{article}
\usepackage{usenix,epsfig,endnotes}
\usepackage{hyperref}
\usepackage{amsmath}
\usepackage{amssymb}
\usepackage{color}
\usepackage{todonotes}
\usepackage{graphicx}
\usepackage{balance}

\usepackage{subcaption}
\begin{document}

\title{\Large \bf On the Effectiveness of Polynomial Realization of Reed-Solomon Codes for Storage Systems}
\author{
  {\rm Kyumars Sheykh Esmaili\thanks{The bulk of this work was done while the author was a Research Fellow at Nanyang Technological University, Singapore.}}\\
    Technicolor Research Lab\\
		Paris, France\\
  kyumars.sheykhesmaili@technicolor.com\\
	\and
  {\rm Anwitaman Datta}\\
    Nanyang Technological University\\
		Singapore\\
  anwitaman@ntu.edu.sg	
}
\maketitle
\thispagestyle{empty}

\subsection*{Abstract}
There are different ways to realize Reed Solomon (RS) codes. While in the storage community, using the \textit{generator matrices} to implement RS codes is more popular, in the coding theory community the \textit{generator polynomials} are typically used to realize RS codes. Prominent exceptions include HDFS-RAID, which uses generator polynomial based erasure codes, and extends the Apache Hadoop's file system.

In this paper we evaluate the performance of an implementation of polynomial realization of Reed-Solomon codes, along with our optimized version of it, against that of a widely-used library (Jerasure) that implements the main matrix realization alternatives.
Our experimental study shows that despite significant performance gains yielded by our optimizations,  the polynomial implementations' performance is constantly inferior to those of matrix realization alternatives in general, and that of Cauchy bit matrices in particular. 

\section{Introduction}\label{sec:introduction}

In the past few years, erasure codes, most prominently Reed Solomon (RS) codes, have been increasingly embraced by distributed storage systems --e.g., Facebook's HDFS-RAID~\cite{HDFS-Facebook}, Microsoft Azure~\cite{EC-Azure}, and Google File System (GFS)~\cite{EC-GFS}-- as an alternative to replication, since they provide high fault-tolerance for low overheads. 

RS codes are defined over Galois Fields of size $2^w$, represented by $GF(2^w)$.  In the RS coding scheme of RS(\textit{k},\textit{m}), an object consisting of $k$ elements (a.k.a symbols) from GF is
encoded into $n = m+k$  blocks (where $n \le 2^w$) in a way that the original object can be recreated from any subset of size $k$ of the $n$ encoded pieces. 


There are two prominent ways to build systematic\footnote{A code is systematic if its encoded output contains all the original $k$ data elements.} Reed Solomon codes. While in the storage community, the \textit{generator matrices} (e.g., Cauchy matrix) have been the dominant realization of RS codes,  in the coding theory community, on the other hand, the \textit{generator polynomials} are the common means to realize RS codes~\cite{wicker}. 
Among the well-known storage systems that uses the polynomial realization is HDFS-RAID~\cite{hdfsraid}, an erasure code-supporting extension of Apache Hadoop's distributed file system (HDFS), developed at Facebook. It has been subsequently used in a number of research prototypes~\cite{XORing,COREBigData,COREUpdate}.

Our goal in this paper is to empirically investigate the effectiveness of the polynomial realization of RS codes and compare its performance against a state-of-the-art implementation of the matrix realization. To this end, we make the following contributions:
\begin{itemize}
	\item describe polynomial realization of RS codes and highlight its distinguishing properties,
	\item build a C mirror for an open source Java implementation of the polynomial realization of RS codes,
	\item explore several techniques to optimize upon the existing polynomial realization, 
	\item conduct a thorough experimental study to investigate the effectiveness of the polynomial realization and compare its performance against Jerasure, an open-source and widely-used library for matrix realization.
\end{itemize}
All our source codes along with a manual can be obtained from an anonymized repository~\cite{RSRepsDL}. We plan to release our implementation as an open-source library.

Our experimental study shows that the polynomial implementations' performance is constantly inferior to those of matrix realization alternatives in general, and that of Cauchy RS codes in particular. This is despite significant performance gains resulted from a range of optimization that we have devised. 

The rest of this paper is organized as follows. We first, in Section~\ref{sec:matrix-rs}, briefly explain the matrix realization of Reed Solomon codes and the two well-known matrix construction methods. Next, a more detailed explanation of the Polynomial realization of RS codes along with its important properties are given in Section~\ref{sec:poly-rs}. Then, after describing the implementation and optimization details in Section~\ref{sec:implementation}, the experimental results are presented in Section~\ref{sec:experiments}. Finally, the paper is concluded in Section~\ref{sec:conclusions}. 

\section{Matrix Realization}\label{sec:matrix-rs}

In this section we first give an overview of the matrix realization of RS codes, and then  briefly  introduce two types of matrices (Vandermonde-based and Cauchy) that are commonly used in storage systems.

\subsection{Overview}\label{sec:matrix-overview}

This realization uses a generator matrix, $\mathbf{G_{n\times k}}$, whose top $k$ rows is an Identity matrix $\mathbf{I_{k\times k}}$.  
One essential property of the generator matrix is that every subset of size $k$ of its rows constitutes an invertible matrix.

To encode $k$ elements of data, its vector is multiplied by $\textbf{G}$, resulting in a codeword composed of the original data vector $d$ and a parity vector $p$ of size $m=n-k$:

	{\small
	$$
	\begin{pmatrix}
		1 & . & . & 0 \\  
		. & . & . & . \\
		. & . & . & . \\
		0 & . & . & 1 \\
		g_{0,0} & . & . & g_{0,k-1} 	\\
		. & . & . & .\\
		. & . & . & .\\
		g_{m-1,0} & . & . & g_{m-1,k-1} 
	 \end{pmatrix}\times
	\begin{pmatrix}
		d_0 \\  
		. \\
		. \\
		d_{k-1}
	 \end{pmatrix}=
	\begin{pmatrix}
		d_0 \\
		. \\
		. \\
		d_{k-1}\\
		p_{0}\\
		. \\
		. \\
		p_{m-1}
	 \end{pmatrix}
	$$
	}

Decoding (i.e., recreating the erased data elements), is performed in 4 steps: (i) rows that correspond to the erased indexes are removed from the generator matrix, (ii) form the surviving rows, $k$ of them are selected to build a matrix of size $k\times k$, (iii) this matrix is inverted, and (iv) multiplying the inverted matrix by the corresponding vector data and parity values will generate the erased elements.

\subsection{Vandermonde-Based Matrices}\label{sec:vandermonde-rs}

The original RS Code~\cite{reed-solomon} is constructed in the following manner. Given a vector of $k$ data elements, the polynomial $P(x)$ is defined as:
$$P(x) = d_0 + d_1x + ... + d_{k-1}x^{k-1}$$
the complete code space $C$ is constructed by choosing $x$ over all possible values in $GF(2^w)$, yielding a system of $2^w$ linear equations, each with $k$ variables:
{\small
$$P(0) = d_0 $$
$$P(\alpha) = d_0 + d_1\alpha + d_2\alpha^2 + ... + d_{k-1}\alpha^{k-1}$$
$$P(\alpha^2) = d_0 + d_1\alpha^2 + d_2\alpha^4 + ... + d_{k-1}\alpha^{2(k-1)}$$
$$...$$
$$P(\alpha^{2^w-1}) = d_0 + d_1\alpha^{2^w-1} + d_2\alpha^{(2^w-1)2} + ... + d_{k-1}\alpha^{(2^w-1)(k-1)}$$
}
Any $k+m$ of the above expressions can be used to construct a RS(k,m) code that can recover from up to $m$ erasures. This is due to the fact that the determinant of the resulting matrices reduces to that of a Vandermonde matrix which is always non-singular and invertible. 

It must be emphasized that none of the matrices built from the above expressions will result in a systematic code, and therefore they are of little use in storage applications. It is, however, possible to transform non-systematic Vandermonde generator matrices into systematic matrices. Details of the transformation method along with numerical examples can be found in~\cite{plank-tutorial, plank-tutorial-new}.

\subsection{Cauchy Matrices}\label{sec:cauchy-rs}

The generator matrix in this case is composed of the identity matrix in the first $k$ rows, and a Cauchy matrix in the remaining $m$ rows. It also has the desired property that all $k \times k$ submatrices are invertible. 
Cauchy Reed Solomon (CRS) coding~\cite{cauchyRS} modifies the scheme in the previous section in two ways.  First, instead of using a Vandermonde matrix, CRS coding employs an $m \times k$ Cauchy matrix. 
which is defined as follows. Let 
$$X = {x_1, . . . , x_m}$$ 
and
$$Y = {y_1, . . . , y_k}$$ 
be such that each $x_i$ and $y_i$ is a distinct element of $GF(2^w)$, and $X \cap Y = \emptyset$. Then the Cauchy matrix defined by $X$ and $Y$ has 
$$\frac{1}{(x_i+y_j)}$$
in element $i$, $j$.

The second modification of CRS is to use projections that convert the operations over $GF(2^w)$ into XORs. 
An important property of these projections is that the multiplication operations --an expensive aspect of the computations-- can be converted into bitwise AND.

It is worth noting that not all Cauchy matrices are equally efficient~\cite{Optimized-CRS}.  There have been some work~\cite{Optimized-CRS,Optimized-CRS2} on generating ``good'' Cauchy matrices, although for a limited set of parameter values.

For numerical examples of CRS encoding and decoding, see~\cite{jerasure,Optimized-CRS}. 

\section{Polynomial Realization}\label{sec:poly-rs} 

Reed-Solomon (RS) codes, in a broader view, are a subset of BCH codes --themselves a class of cyclic error-correcting codes-- whose construction methodology uses \textit{generator polynomials} instead of \textit{generator matrices}. While in the matrix view, codeword is a sequence of \textit{values}, in the polynomial view, it's a sequence of \textit{coefficients}. The codeword space of both views are, however, equivalent through Fourier transformations.

In this section, we first explain the concept of \textit{generator polynomial} and the way that is built, and then describe how encoding and decoding operations are performed for BCH codes.

\subsection{Generator Polynomial}\label{sec:poly-generator}

BCH codes use a generator polynomial, $g(x)$ which consists of $m+1$ factors and its roots are consecutive elements of the Galois Field~\cite{wicker,unb}:

$$g(x)=\sum\limits_{i=0}^{m}g_{i}x^{i} = (x + \alpha^0) (x + \alpha^1) ... (x + \alpha^{m-1})$$

For example\footnote{An extensive set of numerical examples can be found in~\cite{unb}.}, for $RS(k=4,m=3)$ in $GF(2^8)$, where the primitive root is $2$, the generator polynomial will be:
$$
g(x) = (x+2^0)(x+2^1)(x+2^2) = x^3 + 7x^2 + 14x +8
$$

\subsection{Encoding}\label{sec:poly-encoding}

In this realization, the data elements are also represented as $k$ coefficients of a polynomial $d(x)$ of order $k-1$. 
To encode $d$, it is first multiplied by $x^{m}$ and then divided by the generator polynomial, $g(x)$. The coefficients of the \textit{remainder polynomial} $p(x)$  are the output parity elements:
\begin{equation}
d(x) \times x^{m} \equiv p(x) ~~~\bmod~~g(x)
\label{eq:encode}
\end{equation}
or
\begin{equation}
\sum\limits_{i=m}^{m+k-1}d_{i}x^{i}  \equiv   \sum\limits_{i=0}^{m-1}p_{i}x^{i}   \bmod~~  \sum\limits_{i=0}^{m}g_{i}x^{i}
\label{eq:encode-details}
\end{equation}

Given the generator polynomial in the example above, the following data vector:
{\small
$$
48,6,112,70
$$
}
will be encoded as follows
{\scriptsize
$$48x^3 + 6x^4 + 112x^5 + 70x^6 \equiv  243 + 125x + 142x^2~~\bmod~~~ x^3 + 7x^2 + 14x + 8$$
}
hence the parity elements are: 
{\small
$$243,125,142$$
}
One useful property of the encoding operation in the polynomial realization is that the handling of updates is very straightforward. In fact, as shown in~\cite{COREUpdate}, in case of data updates,  encoding the relevant data diff-blocks  will generate the parity diff-blocks. This is in contrast to the matrix realization in which the corresponding coefficients must be extracted from the generator matrix~\cite{zhang2012,peter2012,aguilera2005}.

\subsection{Decoding}\label{sec:poly-decoding}

In the polynomial realization of RS codes, decoding is carried out in two steps:\newline

$\bullet$ \textbf{Step 1}: In this step, \textit{error-evaluator polynomials} are computed. The general specification of these polynomials is as follows:
$$
D(x)  =  p(x) + (\sum\limits_{i=0}^{k-1} {d_{i}}^{\neq 0}x^{i})\times x^{m}
$$
in which ${d_{i}}^{\neq 0}$ denotes the surviving data elements. Given the following equivalency\footnote{This has some commonalities with one of the steps in the Parity-Check Matrix computations explained in~\cite{plank-tutorial-new}.}:
$$
p(x) + d(x) \times x^{m}  ~~~\equiv~~~ 0 ~~~\bmod~~g(x)  
$$
it can be inferred that
$$
D(x)  =  \sum\limits_{i=0}^{k-1} {d_{i}}^{= 0}x^{i}
$$
in which ${d_{i}}^{= 0}$ denotes the erased data elements (to be regenerated). To perform the decoding, a system of equations will be built from as many as $m$ instances of the above formula. 

For example, when the first three  elements of the data vector from the previous example are erased: 
{\small
$$0,0,0,70$$
}
then $D(x)$ is computed for the first three powers of the GF's primitive root: 
{\scriptsize
$$D(\alpha^0) = 243+125(\alpha^0)+142(\alpha^0)^2+70(\alpha^0)^6 = d_0(\alpha^0)^3 + d_1(\alpha^0)^4 + d_2(\alpha^0)^5$$
$$D(\alpha^1) = 243+125(\alpha^1)+142(\alpha^1)^2+70(\alpha^1)^6 = d_0(\alpha^1)^3 + d_1(\alpha^1)^4 + d_2(\alpha^1)^5$$
$$D(\alpha^2) = 243+125(\alpha^2)+142(\alpha^2)^2+70(\alpha^2)^6 = d_0(\alpha^2)^3 + d_1(\alpha^2)^4 + d_2(\alpha^2)^5$$
}

$\bullet$ \textbf{Step 2}: the outcome of Step 1 is a system of equations whose corresponding matrix is of type Vandermonde (note that the size of matrix is not fixed and is determined by the number of erasures). To continue with the example from above:
{\small
$$
\begin{pmatrix}
D(\alpha^0)\\
D(\alpha^1)\\
D(\alpha^2)
\end{pmatrix}
=
\begin{pmatrix}
\alpha^0 & \alpha^0 & \alpha^0 \\
\alpha^3 & \alpha^4 & \alpha^5 \\
\alpha^6 & \alpha^8  & \alpha^{10}
\end{pmatrix}
\times
\begin{pmatrix}
d_0\\
d_1\\
d_2
\end{pmatrix}
$$
}
or
{\small
$$
\begin{pmatrix}
70\\
91\\
171
\end{pmatrix}
=
\begin{pmatrix}
1 & 1 & 1 \\
8 & 16 & 32 \\
64 & 29  & 116
\end{pmatrix}
\times
\begin{pmatrix}
d_0\\
d_1\\
d_2
\end{pmatrix}
$$
}
after solving this system of equations the the erased data blocks are regenerated:
{\small
$$
\begin{pmatrix}
48\\
6\\
112
\end{pmatrix}
=
\begin{pmatrix}
d_0\\
d_1\\
d_2
\end{pmatrix}
$$
}

There are two notable remarks regarding this decoding procedure. 
First, the procedure is symmetric, meaning that it can be applied to the original data vector to generate the parity elements which is basically what the encoding functionality does. An immediate implication of this symmetry is that implementing the decoding procedure is sufficient to provide polynomial RS coding. We have, in fact, exploited this property in our implementation, where we solely focus on optimizing the decoding functionality and use the same method for both encoding and decoding.

Second, while in the matrix-based RS codes, decoding uses exactly $k$ surviving elements (data and parity), the procedure above can use more than $k$ survivor elements and by doing so, speed up the decoding process. In network-critical storage systems, however, the cost of fetching extra data blocks may offset the computational gains.  This tradeoff has been explored in \cite{COREBigData}.

\section{Implementation and Optimization}\label{sec:implementation}

Jerasure~\cite{jerasure} is a widely-used  and open source erasure coding library which implements the matrix realization of RS codes  (both Vandermonde-based and Cauchy variants).  It is written in C and has been shown to be highly efficient~\cite{EC-evaluation}.

For the polynomial realization, to the best of our knowledge, there is no open source implementation in C. There is, however, a Java implementation developed within HDFS-RAID~\cite{hdfsraid} which is built upon $GF(2^8)$. We have ported this implementation to C and improved its performance through a number of optimizations, listed below:\newline

$\bullet$	\textbf{Opt1}: in the first step of the decoding process (Section~\ref{sec:poly-decoding}), HDFS-RAID computes the value of the error-evaluator polynomials ($D(\alpha^i)$'s) in an iterative fashion and independently for each vector of surviving elements. However, since the coefficients of these polynomials (i.e., different powers of the primitive root) are the same for all survivors' vectors, we factor out and pre-compute them beforehand. Then, for each vector of survivors, we just multiply the vector by these common factors.\newline

$\bullet$	\textbf{Opt2}:  in the same stage (Step 1), while the HDFS-RAID implementation always considers all the $m+k$ primitive powers for computing $D(\alpha^i)$'s, our implementation excludes the the erased indexes and hence avoids doing zero-result multiplications.\newline

$\bullet$ \textbf{Opt3}: in the second step of the decoding process, HDFS-RAID employs the Gaussian elimination method to solve the system of equations, once for each vector of polynomial values (computed in Step 1).  But since the same Vandermonde matrix is shared by all vectors, in our implementation we pre-compute and invert this matrix only once. Later, in each iteration we just multiply the inverted matrix by the vector of polynomial values.\newline

$\bullet$ \textbf{Opt4}: while decoding a large number of survivors' vectors, our implementation, uses region-level  multiplication and XORing.  This optimization is inspired by Jerasure and our implementation uses a slightly modified version of one of Jerasure's utility methods. \newline

All of these optimizations are aimed at the decoding functionality, since we exploit the symmetric property of polynomial realization of RS codes and use the same implementation to encode data as well. In terms of effectiveness, based on our development phase tests, the first and third optimizations have the largest impacts. Furthermore, the effectiveness of the first three optimizations grow with increase in the number of erasures.

On top of reducing the computational cost, our optimizations also reduce the memory consumption of the HDFS-RAID through: (i) use arrays of type \textit{char} instead of \textit{int} to represent each $GF(2^8)$ element, and (ii) use of pipelining (e.g., in Opt4) which requires less temporary storage allocations. As a result, the overall memory consumption is decreased significantly (up to 80\%), to a level which is comparable with that of Jerasure.

Lastly, we would like to note that all the source codes (including our optimization codes, HDFS-RAID's implementation of RS coding in C and Java, the Jerasure library, and the experimental utility codes) and a manual can be obtained from~\cite{RSRepsDL}. 

\section{Performance Evaluation}\label{sec:experiments}

In this section, we first explain the important details of our experimental setup and then present the results and analysis.

\subsection{Setup}\label{sec:setup}

In our evaluation study, we have examined the following five methods:
\begin{itemize}
	\item \textbf{OrigCRS}: the original Cauchy RS,
	\item \textbf{GoodCRS}: CRS with ``good'' Cauchy matrices,
	\item \textbf{VanderRS}: the Vandermonde-based RS,
	\item \textbf{PolyRS}: our re-implementation (in C) of HDFS-RAID's implementation of polynomial RS,
	\item \textbf{OptPolyRS}: the optimized version of the above implementation.
\end{itemize}

For the first three methods, we use Jerasure's implementations. In all cases, data and parity elements are defined over $GF(2^8)$, and the multiplication and division tables are pre-computed and maintained in memory.

Similar to~\cite{EC-evaluation}, we focus on the computational cost of encoding and decoding (i.e. recovering from exactly $m$ erasures) operations through measuring  their completion times. Also, in order to minimize the impact of I/O activities, we generate random data and store them in appropriate structures in memory prior to running the encoding and decoding procedures.



The coding scheme used in our experiments is RS(\textbf{k=10},\textbf{m=4}), a popular scheme used by both Facebook~\cite{HDFS-Facebook} and Windows Azure~\cite{EC-Azure}.

For space purposes, here we only report the results of experiments that were run on a  64 bits Debian 7.0 machine  with 4$\times$3.2GHz Xeon Processors and 4GB of RAM. Nonetheless, we would like to note that a large subset of the experiments were replicated on a powerful Windows 7 machine  with a 12$\times$3.20GHz Xeon CPU and 16GB of RAM and a high level of correlation across the two result sets was observed.

Finally, each individual experiment was repeated 3 times and the average values are reported here. Since the variations were very small, we have not included error bars in the graphs.

\begin{figure*}[ht]
				\begin{subfigure}[t]{0.49\textwidth}				
						\centering
						\includegraphics[trim=2cm 5cm 0cm 5cm, width=\textwidth]{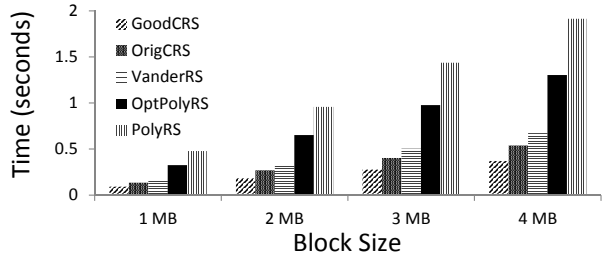}%
						\caption{Encoding Time} 
						\label{fig:block-var-encoding}
				\end{subfigure}			
				\begin{subfigure}[t]{0.49\textwidth}				
						\centering
						\includegraphics[trim=0cm 5cm 0cm 5cm, width=\textwidth]{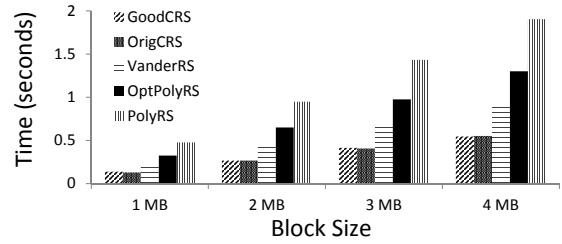}%
						\caption{Decoding Time (for m Erasures)} 
						\label{fig:block-var-decoding}
				\end{subfigure}			
				\vspace{-3mm}
				\caption{Impact of Varying the Block Size in (k=10,m=4)} 
				\label{fig:block-var}				
\end{figure*}			
\begin{figure}[t]
				\centering
				\includegraphics[trim=2cm 5cm 0cm 5cm, width=0.5\textwidth]{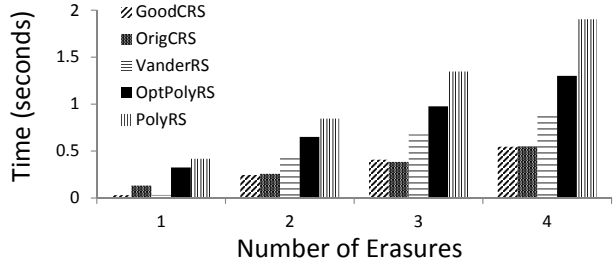}%
				\caption{Decoding Time for Different Erasure Sizes in (k=10,m=4)} 
				\label{fig:erasure-var}
\end{figure}			
\begin{figure*}[ht]
				\begin{subfigure}[t]{0.49\textwidth}				
					\centering
					\includegraphics[trim=2cm 5cm 0cm 5cm, width=\textwidth]{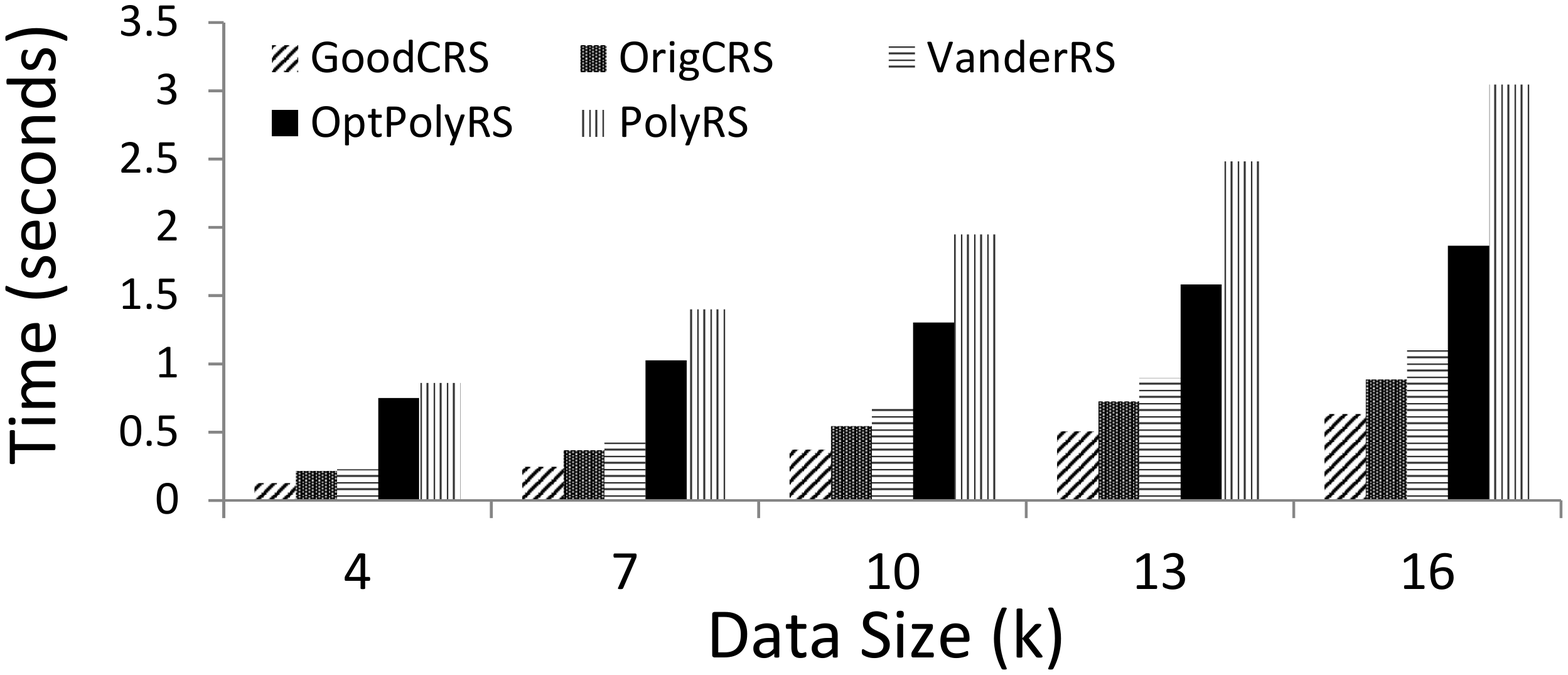}%
					\caption{Encoding Time}%
					\label{fig:k-var-encoding}%
				\end{subfigure}			
				\begin{subfigure}[t]{0.49\textwidth}				
					\centering
					\includegraphics[trim=0cm 5cm 0cm 5cm, width=\textwidth]{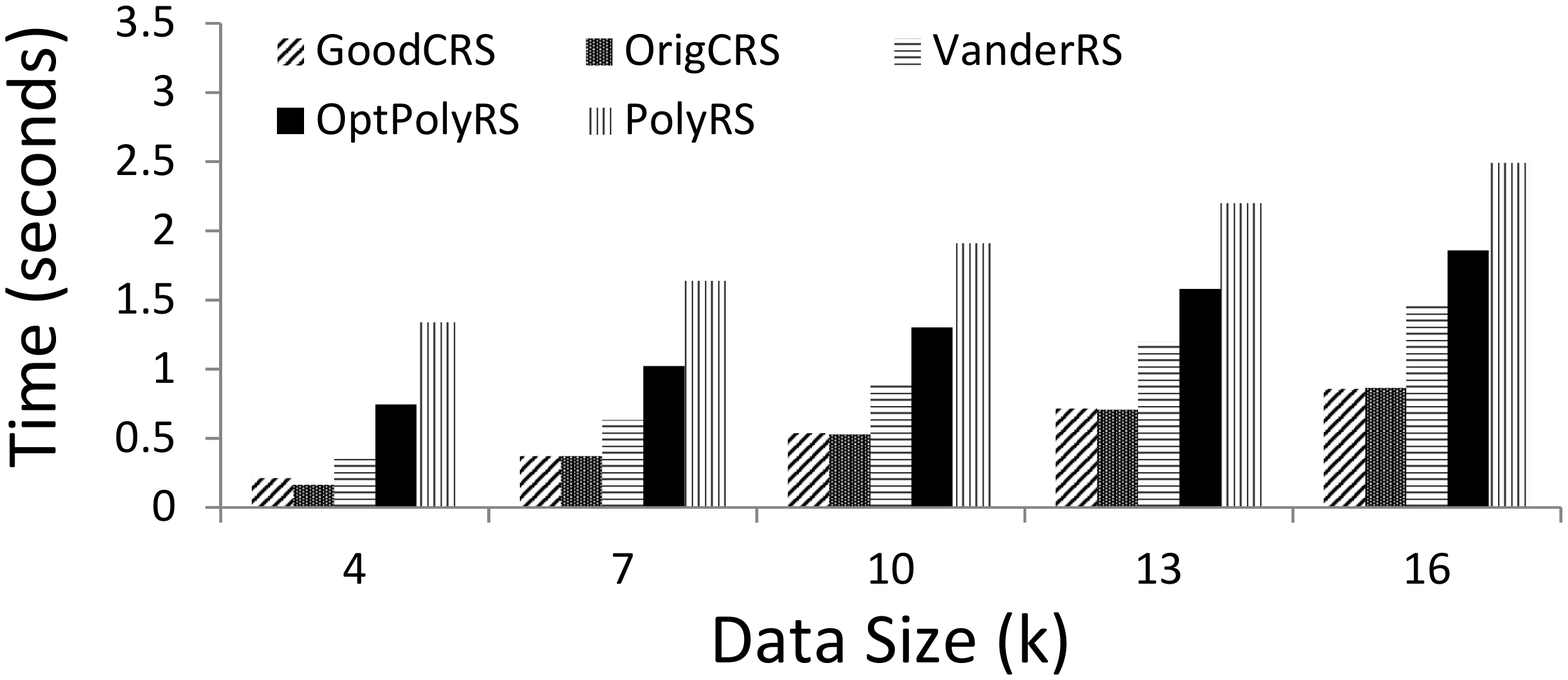}%
					\caption{Decoding Time}%
					\label{fig:k-var-decoding}%
				\end{subfigure}
				\vspace{-3mm}
				\caption{Impact of Varying the Data Size for m=4} 
				\label{fig:k-var}
				
				\begin{subfigure}[t]{0.49\textwidth}				
					\centering
					\includegraphics[trim=2cm 5cm 0cm 4cm, width=\textwidth]{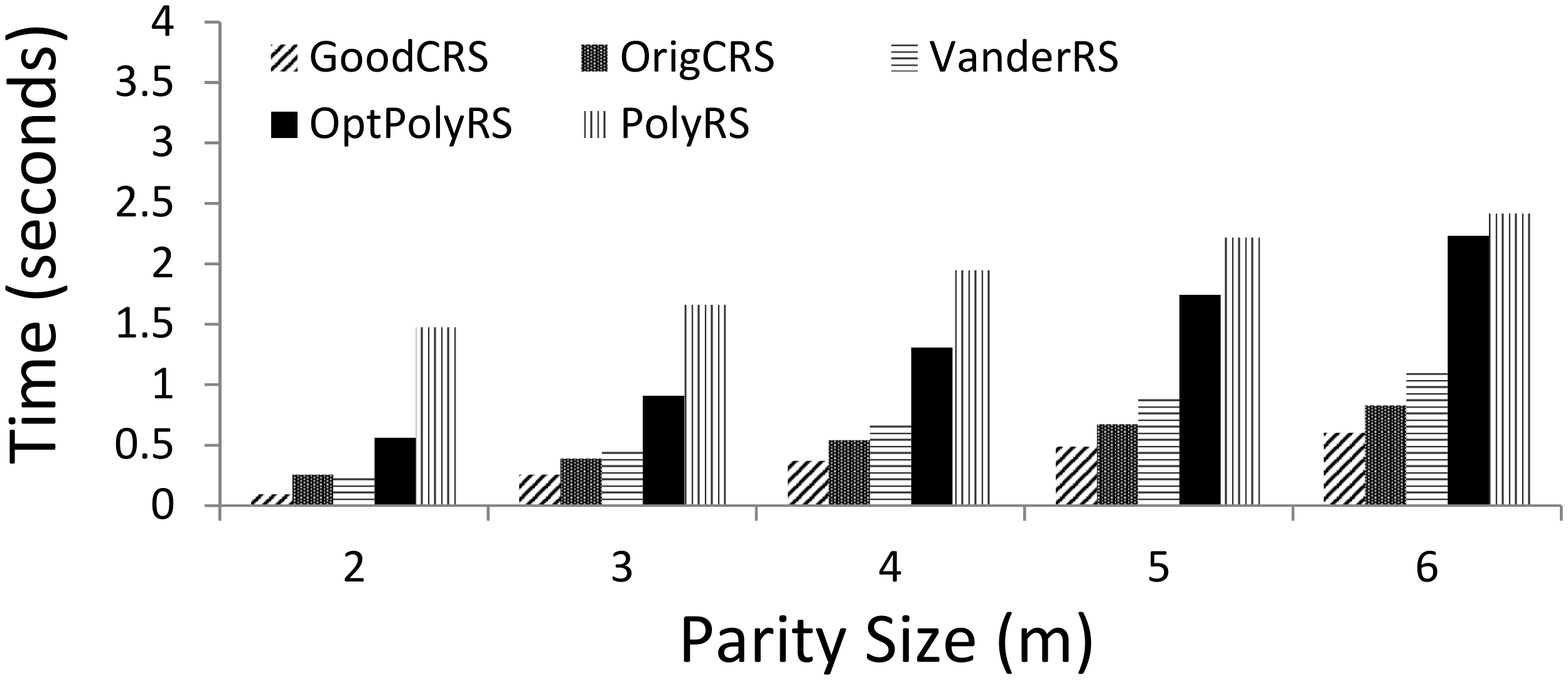}%
					\caption{Encoding Time}%
					\label{fig:m-var-encoding}%
				\end{subfigure}			
				\begin{subfigure}[t]{0.49\textwidth}				
					\centering
					\includegraphics[trim=0cm 5cm 0cm 4cm, width=\textwidth]{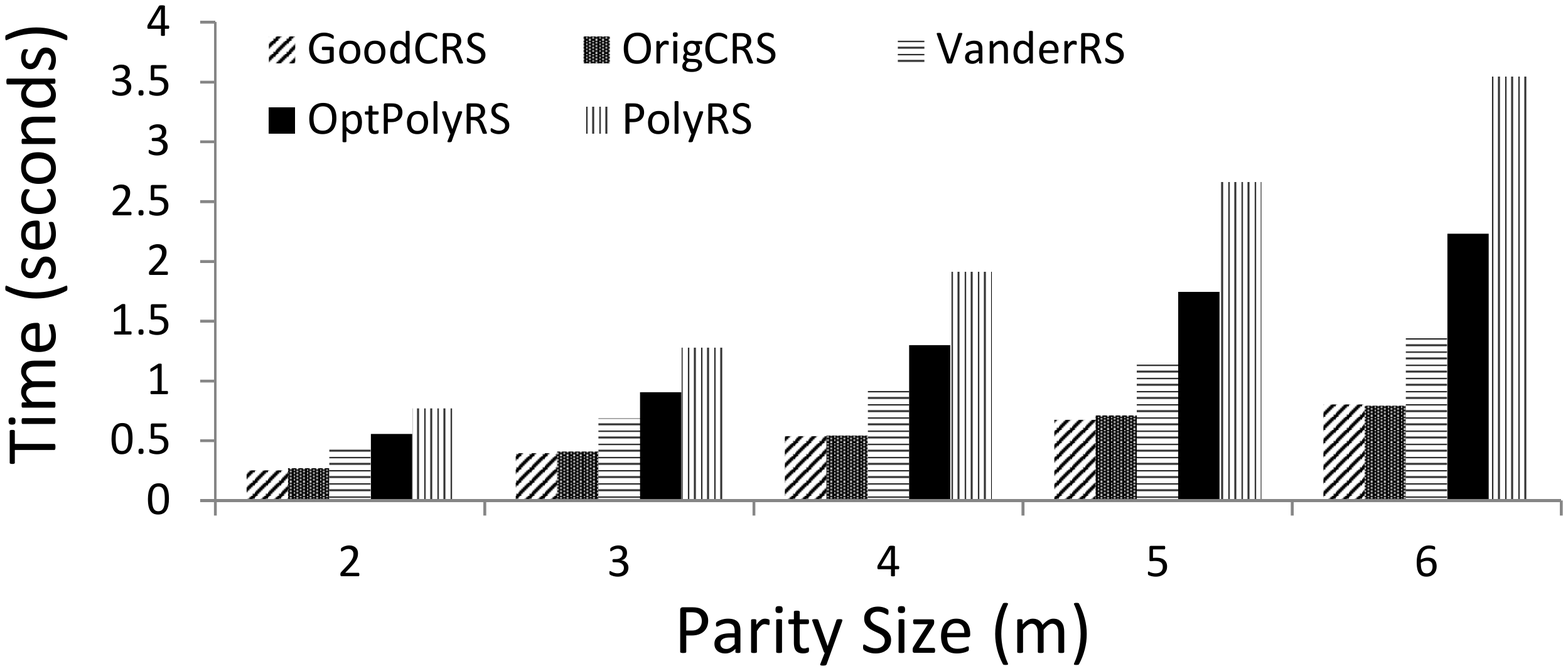}%
					\caption{Decoding Time}%
					\label{fig:m-var-decoding}%
				\end{subfigure}
				\vspace{-3mm}
				\caption{Impact of Varying the Parity Size for k=10 } 
				\label{fig:m-var}			
\end{figure*}

\subsection{Results and Analysis}\label{sec:results}

In our experiments, we varied a number of crucial parameters:\newline
$\bullet$ \textbf{Block Size}. A block is a coarser-grain collection of GF elements (data or parity). We vary the block size parameter from 1MB to 4MB. In our scheme of (10,4), this means that total size of data and parity changes from 14MB to 56MB. The results of this experiment are depicted in in Figure~\ref{fig:block-var}.  For all the remaining experiments, the block size is set to be 4MB.\newline
$\bullet$ \textbf{Erasure Size}. In this experiment we vary the erasure size from its minimum, 1, to its maximum, 4. The results are shown in Figure~\ref{fig:erasure-var}. In all other experiments the erasure size is maximum.\newline
$\bullet$ \textbf{Coding Parameters}. We  vary both data size, $k$, and parity size, $m$. Note that changing either of two parameters, changes the storage overhead (i.e., the $m/k$ ratio) of coding scheme, although in opposing  directions. The results are summarized in Figure~\ref{fig:k-var} and Figure~\ref{fig:m-var}, respectively.

Based on the above results, here are the notable patterns:
\begin{itemize}
	\item Matrix-based implementations consistently --in all parameter combinations and for both encoding and decoding-- outperform the polynomial ones. Furthermore, they generally have lower growth rates (slopes) as well.		
	\item In real world scenarios, single erasures (per stripe) are by far the most common type of failures in storage systems~\cite{HDFS-Facebook}. As such, based on the results presented in Figure~\ref{fig:erasure-var}, matrix-based realizations have a significant advantage (up to 10 times faster). In network-critical configurations, the differences will be even higher (as explained in Section~\ref{sec:implementation}).	
	\item The decoding of matrix methods are more effective for higher storage overheads. 
	\item The optimization gains in OptPolyRS increase with the number of erasures (inline with our development phase tests, as mentioned in Section~\ref{sec:implementation}).
	\item Data encoding in PolyRS is considerably slow for low storage overheads e.g., it requires more than \%250 of our optimized version's time in (10,2). The gap, however,  narrows as the storage overhead grows (around \%10 in (10,6)). 	
	\item GoodCRS is more effective (compare to OrigCRS) in encoding than in decoding.
\end{itemize}

\section{Conclusions and Future Work}\label{sec:conclusions}

We evaluated the performance of an implementation of polynomial-based Reed-Solomon codes against that of a state-of-the-art implementation of two main matrix-based alternatives. 
Based on our experimental study, the polynomial implementation's performance is constantly inferior to those of matrix alternatives in general, and that of Cauchy Reed Solomon in particular. This is despite significant performance gains resulted from a range of optimization that we have devised. 

One important conclusion to draw from these results is that HDFS-RAID's RS coding performance can be greatly improved, by either adopting some of the optimizations described in this paper, or by using Cauchy matrices RS codes.

We see three directions to extend the work reported here. Firstly, since one of the main factor behind CRS's high efficiency is its use of bit-matrix and multiplication-free computations, and given the high number of multiplications in the polynomial realization, it would be interesting to see what impact the adoption of a bit-matrices will have on it.

Secondly, a recent paper~\cite{Plank13} has demonstrated that it is possible to multiply regions of bytes by constants in a Galois Field very fast. It's not quite as fast as XOR, but the speed is limited by how quickly you can populate the L3 cache. Another way to extend our current work is to integrate this new multiplication method and measure its impact.

Lastly, as the methodology and proofs in~\cite{hall} show, a generator polynomial can be transformed into a generator matrix. Examining the effectiveness of such matrices for RS coding can be another avenue for future work.

\balance


\end{document}